\documentclass[%
superscriptaddress,
 reprint,
 amsmath,amssymb,
 aps,
 prb,
]{revtex4-2}

\usepackage{graphicx}
 \usepackage[normalem]{ulem}
\usepackage[utf8]{inputenc}
\usepackage[T1]{fontenc}
\usepackage{etoolbox}
\usepackage{float}
\usepackage{xcolor}
\usepackage{upgreek}

\makeatletter
\def\@email#1#2{%
 \endgroup
 \patchcmd{\titleblock@produce}
  {\frontmatter@RRAPformat}
  {\frontmatter@RRAPformat{\produce@RRAP{*#1\href{mailto:#2}{#2}}}\frontmatter@RRAPformat}
  {}{}
}%
\makeatother

\newcommand{\blue}{\textcolor{black}}

\begin{document}


\title{Role of quantum confinement in semiconductor-superconductor core-shell nanowires}

\author{Tudor Gabriel Dumitru}
\affiliation{Department of Engineering, Reykjavik University, Menntavegur 1, IS-102 Reykjavik, Iceland.}

\author{Anna Sitek}
\affiliation{Institute of Theoretical Physics,
	Wroclaw University of Science and Technology,
	Wybrze{\.z}e Wyspia{\'n}skiego 27, 50-370 Wroclaw, Poland.}

\author{Gunnar Thorgilsson}
\affiliation{Department of Engineering, Reykjavik University, Menntavegur 1, IS-102 Reykjavik, Iceland.}
    
\author{Sigurdur I.\ Erlingsson}
\affiliation{Department of Engineering, Reykjavik University, Menntavegur 1, IS-102 Reykjavik, Iceland.}

\author{Tudor Dan Stanescu}
\affiliation{Department of Physics and Astronomy, West Virginia University, Morgantown, WV
26506}

\author{Andrei Manolescu}
\affiliation{Department of Engineering, Reykjavik University, Menntavegur 1, IS-102 Reykjavik, Iceland.}

\begin{abstract}
This work is motivated by the experimentally observed coherence of the supercurrent in semiconductor nanowires covered by a half-shell metallic superconductor, which leads to flux dependent supercurrent oscillations with period $h/2e$, as expected for a tubular superconductor, i.e. Little-Parks oscillations.  We perform microscopic model calculations and compare the results for full and half metallic shells.  We use an effective Hamiltonian derived from the Green's function of the proximitized semiconductor nanowire, where the presence of the superconductor is represented by a self energy. Furthermore, we incorporate the electrostatic band-bending at the metal-semiconductor interface as a rectangular narrow quantum well on the semiconductor side. The properties of the eigenstates of the effective Hamiltonian are determined by the spatial profile of the corresponding transverse modes in the normal state. For half-shell wires, transverse modes with high-enough energy expand outside the interface quantum well and generate eigenstates with mixed electron-hole character that surround the entire circumference of the nanowire, similar to eigenstates of the full-shell system. We identify these states as being responsible for the observed Little-Parks effect. 
\end{abstract}

\maketitle

\section{Introduction \label{sec:intro}}
Core-shell nanowire heterostructures are fundamental building blocks for a diverse range of nanoelectronic and optoelectronic devices, valued for their highly tunable properties and synthetic feasibility. Recent advances in fabrication have enabled the development of more complex core-multiple-shell architectures, which consist of a nanowire core encased in several concentric layers of different materials. These intricate structures offer additional degrees of freedom for engineering electronic properties, such as band alignment and charge carrier localization. 

The internal geometry of the nanowire is, in particular, important. Grown via a bottom-up approach, experimental III-V semiconductor nanowires are typically prismatic rather than cylindrical, with a cross-section that reflects the underlying crystal structure. The most common is the hexagonal shape of the cross section \cite{Blomers13, Rieger12, Fickenscher13, Funk13, Haas13, jadczak14,Erhard15, Shi15, Boland15, Sonner19, Ra20, Cereti22, Zhang2024, AyusoPerez2024, Sun2025, Peeters25}. The geometry of a prismatic shell is of significant physical interest due to the possible localization of carriers at the corners or on the sides of the shell \cite{Ferrari09, Bertoni11}.  Other configurations, such as square \cite{Fan06, Shtrikman09,Hu11}, triangular \cite{Dong09, Qian04, Qian05, Baird09, Yuan15, Goransson19}, or even dodecagonal \cite{Rieger15} cross-sections, have also been reported. 

The core-shell structure has a particularly compelling application in the creation of hybrid semiconductor-superconductor (semi-super) devices \cite{Benito20, Paya25a, Paya25b}, which have complex properties and are a leading platform for realizing topological qubits \cite{Hays21,Aguado20}. In these devices, a superconducting material induces a pairing potential in an adjacent semiconductor via the proximity effect, wherein Cooper pairs from the superconductor penetrate the semiconductor and diffuse \cite{Schapers_book}. This phenomenon has been the subject of intense investigation, largely driven by the search for Majorana zero modes, which are predicted to emerge at the ends of such nanowires in the presence on an applied magnetic field \cite{Lutchyn10,Oreg10,Prada20,Stanescu17a,DSarma23,kouwenhoven24}. 

The induced superconductivity that emerges in hybrid core-shell semiconductor-superconductor structures has been probed in recent transport experiments.
 In particular, the current measured along a semiconductor nanowire made of InAs completely surrounded by a thin superconductor shell made of Al, in the presence of a longitudinal magnetic field, has shown flux-periodic oscillations with period $h/2e$ \cite{Vaitiekenas20}, which corresponds to the Little-Parks effect originally observed in tubular superconductors. This implies that the electrons hosted by the semiconductor accumulate in the proximity of the semiconductor-metal interface, forming a thin superconducting shell inside the InAs nanowire. 
 
 Similar experiments were performed on semiconductor nanowires cover by only a half metallic superconductor (Al) shell. Surprisingly, the critical current measured in the semiconductor has shown  again flux-periodic Little-Parks oscillations \cite{Stampfer22,Zellekens25}, as if it was flowing through a complete tubular superconductor. In the normal state, a tubular distribution of the charge carriers inside the semiconductor nanowire was obtained either as a surface accumulation layer (in InAs), due to the pinning of the Fermi level \cite{Stampfer22}, or as a result of the core-shell GaAs-InAs radial geometry \cite{Zellekens25}.  But the surprise was that the induced superconducting phase was coherent around the entire circumference of the semiconductor shell, even in regions far away from the interface, i.e. not in direct contact with the superconductor. This raises the fundamental question of how far away from the interface is superconductivity induced inside of the semiconductor nanowire.  


To describe the underlying physics of these complex devices and fully capture the intricate phenomena characterizing the hybridized structures, traditional phenomenological models based on Andreev electron-hole retro-reflections prove insufficient. Instead, we propose a computational quantum mechanical approach based on a microscopic Bogoliubov-de Gennes (BdG) model of a semiconductor nanowire in contact with the metallic superconductor.  We construct the effective BdG Hamiltonian starting with the Green's function of the hybrid system, in which the contribution from the parent superconductor is incorporated as an interface self-energy term.  

The band bending effects associated with the work function difference across the interface are modeled using a simple quantum well (QW) potential, as shown in Fig.~\ref{fig:sample}. We note that, although the actual electrostatic potential inside the semiconductor has a more complicated spatial dependence, this simple QW model accurately captures the key physics associated with the electrons inside the semiconductor being attracted toward the semiconductor-superconductor interface. We also note that the presence of a surface accumulation layer (e.g. in InAs) can be modeled using a similar QW potential on the free (i.e. unproximitized) surface of the semiconductor nanowire \cite{Schuwalow21}. 

We diagonalize the effective BdG Hamiltonian numerically and systematically compare the results corresponding to a full and a half metallic shell surrounding the semiconductor nanowire.   
Our main goal is to establish if the half-shell geometry is consistent with the existence of BdG states that (i) have a mixed electron-hole character, i.e., reflect the presence of induced superconductivity, and (ii) fully surround the core of the nanowire. This type of states occur naturally in full-shell hybrid nanowires and are responsible for the Little-Parks effect observed in these systems.
We discuss how the ``tubular'' superconducting states are affected by factors such as the interface potential, the semiconductor thickness, and the wire length.

The remainder of the paper is organized as follows:  In Section \ref{sec:model} we describe the physical model and the computational methods.  The results are presented in Section III, which is organized in three subsections that discuss (A) the transverse modes, (B) the properties of infinitely-long nanowires, and (C) the low-energy physics of finite length systems.  The conclusions are collected in Section IV.

\section{The model and the computational approach\label{sec:model}}

\begin{figure}
\includegraphics[trim=0 100 0 40, clip, scale=0.39]{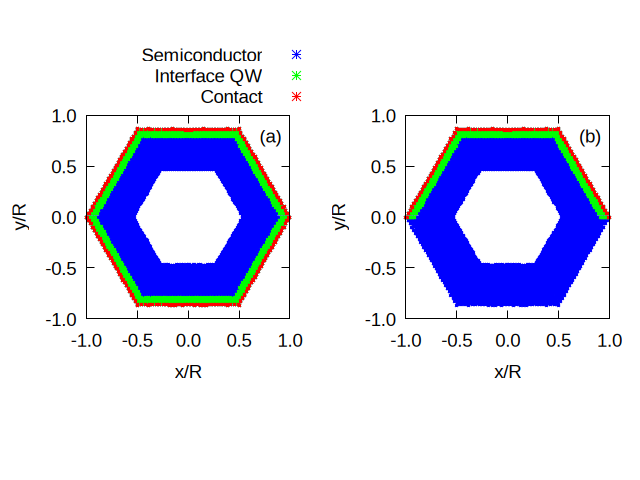} 
\caption{Cross section of the hybrid core-shell nanowires with (a) full-shell and (b) half-shell superconductor layers. The white region in the center is the (insulating) semiconductor core, while the blue and green regions represent the (conducting) semiconductor shell, where charge carriers are present. The green zone indicates the QW potential created at the metal-semiconductor contact. The red line corresponds to the
self-energy which in our model represents the interface with the superconducting layer. 
}
    \label{fig:sample}
\end{figure}

A three-dimensional hexagonal core-shell nanowire is modeled using cylindrical coordinates, with the cross section of the nanowire in the $x-y$ plane, and the $z$ direction along the longitudinal axis. The semiconductor nanowire is covered by a superconducting layer, which is either a full shell, or a half shell toping three facets of the hexagonal profile, illustrated in Fig.~\ref{fig:sample}. While the superconducting metal is not explicitly included in the model, its effects are incorporated through (i) a potential well within a relatively narrow semiconductor region near the semiconductor-metal interface, which accounts for the electrostatic effects generated by the nonzero work function difference \cite{Woods19}, and (ii) a self-energy interface term, which accounts for the superconducting proximity effects \cite{Stanescu22}. 
The semiconductor material is a core-shell nanowire, with a conducting shell (n-doped) surrounding an insulating core. The corresponding cross section is illustrated in Fig.~\ref{fig:sample}.  In the $z$ direction, the nanowire is assumed to be either infinite, or finite (of length $L_z$).

The first step is to compute the transverse modes, i.e. the eigenstates corresponding to a cross section of the (normal state) wire, which is modeled using a polar grid with imposed polygonal boundaries \cite{Daday11}. The points inside these boundaries are selected, while the ones outside are discarded, leaving only the desired hexagonal shape \cite{Sitek15}.  The model includes a QW potential on the sites situated in close proximity to the metal-semiconductor interface, i.e the area doped with electrons due to the presence of the metallic layer (green shading in Fig.~\ref{fig:sample}). 
An ab initio density-functional method was used to extract the Hartree potential of the heterostructure, resulting in a complex gradual profile with a well on the metal side and a barrier on the semiconductor side \cite{Dumitru26a}. For simplicity, here we approximate this complex profile using only a thin rectangular potential well at the metal-semiconductor interface.  This simplified QW potential incorporates the electrostatic effects generated by the work function difference at the semiconductor-metal interface.  
The values of the parameters used in our model are listed at the end of this section.

We diagonalize numerically  the transverse Hamiltonian on our polar lattice and find the transverse wave functions $|a\rangle$ and the associated energies $\epsilon_a$, where $a=1,2,3,\ldots$ labels states in order of increasing energy, $\epsilon_1\leq \epsilon_2\leq\dots$ .  
In the $z$ direction the wave functions are either plane waves with wave vector $k$, or sinusoidal modes with $n=1,2,3, \dots$, for infinite or finite length $L_z$, respectively. Explicitly, we have
\begin{equation}
|k\rangle = \frac{1}{\sqrt{L_z}} e^{ikz} \, , \quad {\rm or} \quad  
|n\rangle = \sqrt{\frac{2}{L_z}} \sin \left(\frac{n\pi z}{L_z} \right)  \,.
\label{eq:sine}
\end{equation}
 The Hamiltonian of the nanowire in the normal state, $H_w$, has eigenstates $|m\rangle$ given by the solutions of the eigenvalue equation $H_w|m\rangle=E_m|m\rangle$. More specifically, we have 
$|m\rangle=|ak\sigma\rangle$ or $|m\rangle=|an\sigma\rangle$, for infinite and finite $L_z$, respectively,  
where $\sigma$ is the spin index, and the corresponding eigenvalues are
\begin{equation}
E_m=\epsilon_a+\frac{\hbar^2k^2}{2m_{\rm eff}} \ ,
\hspace{3mm} {\rm or} \hspace{3mm} 
E_m=\epsilon_a+\frac{\hbar^2\pi^2n^2}{2m_{\rm eff}L_z^2} \ .
\end{equation}
Here, $m_{\rm eff}$ denotes the effective mass of the semiconductor and we used the tensor product notation 
$|a k \sigma\rangle \equiv |a\rangle\otimes |k\rangle\otimes|\sigma\rangle$.

The next step is to incorporate the proximity-induced superconductivity. We assume that the thin film proximitizing the semiconductor nanowire (as either a full or a half shell), is a superconductor with gap parameter $\Delta$. After integrating out the superconducting degrees of freedom, the effect of the metallic superconductor on the semiconductor material is taken into account via the self energy \cite{Stanescu13,Stanescu17b}
\begin{equation}
\begin{split}
& \Sigma(E;{\bf r}) = -{\boldsymbol\gamma}({\bf r}) \ \times \\ 
& \left( 
\frac{E}{\sqrt{\Delta^2 - E^2}} \, {\bf 1} \otimes {\bf 1}
+ \frac{\Delta}{\sqrt{\Delta^2 - E^2}} \, \sigma_y \otimes \tau_y \right) \ ,
\end{split}
\label{eq:se}
\end{equation}
where ${\boldsymbol\gamma}({\bf r})$ represents the effective semiconductor-superconductor coupling strength, $\sigma_y$ and $\tau_y$ are the Pauli matrices in the spin and Nambu spaces, respectively, and the symbol {\bf 1} designates the unit matrix in these spaces.  The semiconductor nanowire with the proximity induced superconductivity is described by the Green's function   
\begin{equation}
G(E) = \frac{1}{E-\widetilde{H}_w-\Sigma} \, ,
\label{eq:gf}
\end{equation}
where $\widetilde{H}_w$ is the Hamiltonian of the nanowire expanded to the Nambu space. The density of states (DOS) can be obtained as
\begin{equation}
\rho(E) = -\frac{1}{\pi} \text{Im} \ \text{Tr} \ G^- ,
\label{eq:dos}
\end{equation}
where the trace is taken over all spatial ($a$ and $k$ or $n$) and all spin and Nambu labels, and the minus upper label designates the retarded Green's function.

The superconducting gap induced in the semiconductor shell is expected to be smaller (possibly much smaller) than the parent gap $\Delta$.  Therefore, if we are interested in low-energie states with $E \ll \Delta$, we can consider $\Delta^2 - E^2\approx \Delta^2$, and the self energy (\ref{eq:se}) becomes 
\begin{equation}
\Sigma(E;{\bf r}) \approx -\boldsymbol\gamma({\bf r})
\left( 
\frac{E}{\Delta} \, {\bf 1} \otimes {\bf 1}
+ \sigma_y \otimes \tau_y \right) \ ,
\label{eq:ses}
\end{equation}
which corresponds to the so-called static approximation 
\cite{Stanescu13}.  Within this approximation, since the self-energy becomes a linear function of energy, we can solve the eigenvalue problem $\left[\widetilde{H}_w+\Sigma(E;{\bf r})\right]\psi_\alpha=E_\alpha\psi_\alpha$. 
Explicitly, we introduce the operator $Z^{-1}=\left[1+\boldsymbol\gamma({\bf r}) / \Delta\right] {\bf 1} \otimes {\bf 1}$ and rewrite the eigenvalue problem as
\begin{equation}
H_{\rm eff}\Phi_{\alpha}=E_{\alpha} \Phi_{\alpha}\ ,
\label{eq:heff}
\end{equation}
where $H_{\rm eff}=Z^{1/2} \widetilde{H}_w Z^{1/2}$ is the effective BdG Hamiltonian and $\Phi_{\alpha}=Z^{-1/2}\psi_\alpha$. Note that, assuming $\langle \Phi_\alpha|\Phi_\alpha\rangle=1$, the eigenstates $\psi_\alpha$ are not normalized to $1$. The ``missing'' spectral weight, $1-\langle \psi_\alpha|\psi_\alpha\rangle$, represents the contribution corresponding to the component of the state $\alpha$ localized inside the parent superconductor, which is not explicitly described by the effective model. The spectrum of $H_{\rm eff}$ provides an excellent approximation of the eigen energies of the hybrid system up to $E_\alpha\leq \Delta/2$, while the corresponding eigenstates give information regarding the spatial dependence of the actual eigenstates inside the semiconductor wire and regarding their electron-hole content. 

We computed the eigenstates of the Hamiltonian $H_{\rm eff}$ within a two-step process. First, we obtained the transverse modes on a polar lattice with 13000 points. Second, we used the first 12 transverse modes and 70 longitudinal modes given by Eq.~(\ref{eq:sine}) to construct the $H_w$ matrix, and then, after expanding it to the Nambu space ($H_w\rightarrow \widetilde{H}_w$) we construct the effective Hamiltonian matrix, $H_{\rm eff}$.

In the next section we will present the results of our calculations based on the following list of parameters: The semiconductor shell, shown with blue color in Fig.~\ref{fig:sample}, has a cross sectional external radius $R_{\rm ext}=100$ nm (measured between the center and an outer corner), with the side thickness of 40 nm (that will be reduced later for comparison), and it is assumed to be made of InAs, with electron effective mass $m_{\rm eff}=0.023m_0$. The QW created at the interface with the metallic full or half shell, shown in green in Fig.~\ref{fig:sample} (a) and (b), respectively, is 50 meV deep and 8 nm wide.  We use the superconducting gap parameter of the metal $\Delta = 0.33$ meV (a little larger than what one expects from a thin Aluminum shell) and the coupling function $\boldsymbol\gamma({\bf r})=\gamma=83$ meV at the contact region with the semiconductor, shown with red color in Fig.~\ref{fig:sample}, and zero in the rest of the volume.  This parameter was established numerically, as will be shown in the next section.  The length of the nanowire will be either infinite, or in the range 1000-5000 nm.

\section{Results \label{sec:res}}

\subsection{The transverse modes}

\begin{figure}
\centering
\includegraphics[trim=0 0 0 60, clip, scale=0.39]{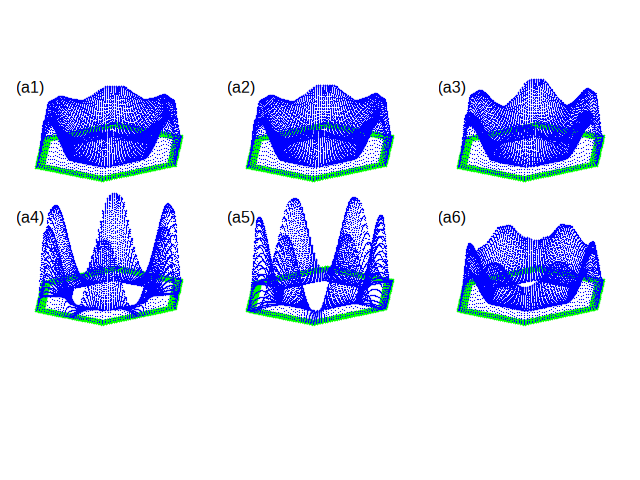} \\[-20mm]
\includegraphics[trim=0 0 0 30, clip, scale=0.39]{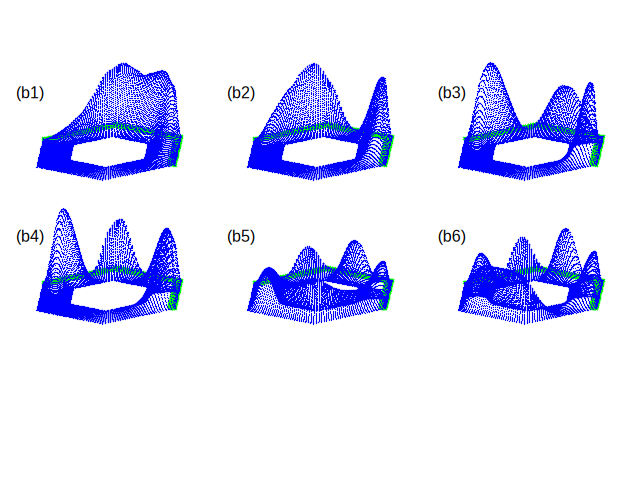}  \\[-20mm]
\caption{(a1-a6) Wave functions corresponding to the lowest energy transverse modes of a full shell nanowire, in increasing energy. The green shading indicates the QW potential. Modes (a2), (a3), and (a6) are double degenerated because of the hexagonal symmetry. Note that the electrons are shifted towards the external boundaries, where the QW potential is located. 
(b1-b6) Transverse wave functions for the half shell case, with metal-semiconductor contact on three facets. 
Note that the lowest four energy modes (b1-b4) are localized near the metal-semiconductor interface, while the 5-th and 6-th modes extend throughout the entire semiconductor shell, i.e. they are tubular-type states similar to those in the (a) panels.}
\label{fig:tm}
\end{figure}

First, we calculate the wave functions of the transverse modes supported by the semiconductor wire in the normal state. Results corresponding to the low energy modes are shown as localization probabilities in Fig.~\ref{fig:tm}.  The series of figures a(1-6) and b(1-6) show the first modes, in increasing energy order, for the full and half metal shell, respectively. We notice prominent peaks in the corner regions of the hexagonal cross section. In the full shell case, the localization follows the symmetry of the hexagonal cross section of the nanowire, and the QW (shown in green) pulls the distribution a bit outwards. Due to the discrete rotational symmetry, some transverse modes are degenerated with respect to the orbital motion.  The order of this orbital degeneracy follows a periodic sequence with period 1,2,2,1, or 2,4,4,2, if the spin is included \cite{Sitek15}. The modes shown here are (on the energy scale): mode 1 - the ground state (a1), then modes 2 and 3 (a2), modes 4 and 5 (a3), mode 6 (a4), mode 7 (a5), and modes 8 and 9 (a6). 
All states are spin degenerated, but the spin degree of freedom was not explicitly included in the transverse Hamiltonian.  

No orbital degeneracy occurs in the half shell case, as the hexagonal rotation symmetry is explicitly broken by the QW potential. The lowest energy modes are bound by the QW potential within half of the wire cross section. However, the 3-rd and 4-th modes have visible tails outside the well, while the 5-th and 6-th modes become distributed over the entire perimeter of the cross section. 
Thus, although the effective electrostatic potential induced by the contact with the metal localizes the lowest energy transverse modes in the vicinity of the metal-semiconductor interface, the modes with high-enough energy extend throughout the whole shell, i.e., they are tubular-type transverse states. This behavior is generic and our simple modeling of the effective electrostatic potential using a QW does not introduce qualitative limitations. The key point is that core-shell semiconductor nanowires covered by a half-shell superconducting film can support tubular states, if the chemical potential is high-enough. The critical next question is to what degree these state carry proximity-induced superconductivity.

\subsection{Nanowires with infinite length}

\begin{figure}
\centering
\includegraphics[trim=0 0 0 120, clip, scale=0.39]{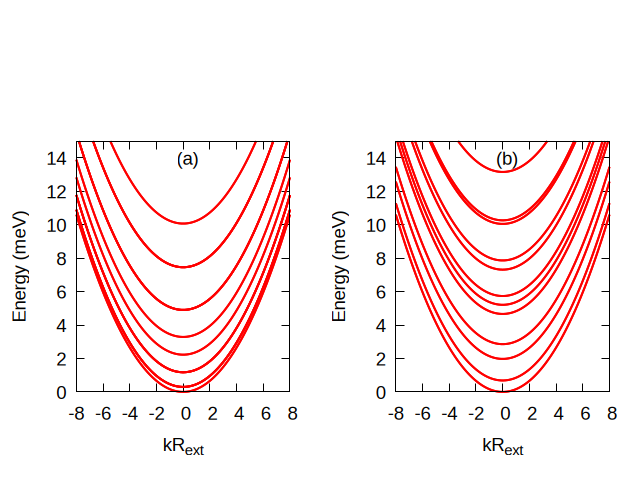}  \\[-3mm]
\caption{Energy spectra for the nanowire with infinite length in the normal state. The transverse energies correspond to the bottom of each parabola, relatively to the ground state which is set at zero energy. (a) Full shell case, where the degeneracy sequence is 2, 4, 4, 2 due to orbital and spin degeneracies \cite{Sitek15}. (b) Half shell case, where the orbital degeneracy is lifted, and all parabolas are only 2-fold spin degenerate. 
}
\label{fig:ien}
\end{figure}

Having determined the transverse modes, we can easily obtain the total states and the corresponding energies for the infinitely-long nanowire by exploiting the translation invariance of the system in the $z$ direction. The energy spectrum has parabolic dispersion in $k$, as shown in 
Fig.~\ref{fig:ien}, where the zero of the energy was chosen at the single-particle ground state. The  transverse energies $\epsilon_a$, $a=1,2,3,\ldots$, correspond to the bottom of each parabola, where $k=0$. In Fig. \ref{fig:ien}(a), some of the parabolas, e.g., second, third, sixth, etc., as discussed above, have orbital degeneracy and, in addition, all the states in Fig.~\ref{fig:ien} are doubly spin degenerate.
Note that the highest two modes shown in Fig.~\ref{fig:tm}, for the full shell case, i.e. (a5) and (a6), correspond to the fifth and sixth parabolas in Fig \ref{fig:ien}(a) with energies (at $k=0$) 3.3 meV and 4.9 meV, respectively. Whereas for the half shell case the modes (b5) and (b6) have higher energies, 4.7 meV and 5.2 meV, respectively, (at $k=0$) due to the QW having a shorter length, i.e. only half of the wire circumference. 
We note that for the half metallic shell, in order to have tubular states that surround the entire nanowire volume, the chemical potential should be larger than $\sim 5$ meV, so that some of the (b5) and (b6) modes are occupied. 

\begin{figure}
\centering
\includegraphics[scale=0.38]{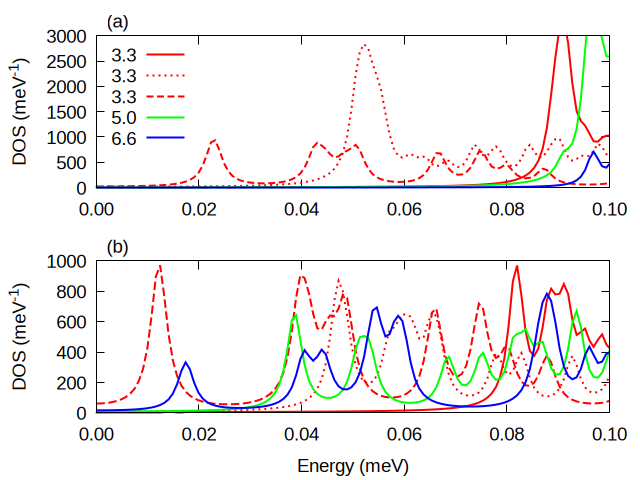}  \\[-2mm]
\caption{Density of states for (a) the full shell model and (b) the half shell model with several sets of parameters. The values of the chemical potential are  explicitly given in panel (a) in meV units.
Full lines correspond to a superconductor-semiconductor coupling constant $\gamma=83$ meV and a QW potential of $-50~$meV. The dotted lines represent the DOS of hybrid systems with reduced semiconductor-superconductor coupling, $\gamma=41.5$ meV, while the dashed lines correspond to a vanishing QW potential.
}
    \label{fig:dos}
\end{figure}

In order to explore the superconductivity induced in the semiconductor material, we calculated the matrix elements of the Green's function using  Eqs.~(\ref{eq:se}-\ref{eq:gf}).  Since the self energy is real for $E<\Delta$, when inverting the Green's function's matrix we added a small imaginary constant of 0.002 meV to $\Sigma$ to avoid the singularities.  The DOS, Eq.~(\ref{eq:dos}), is shown in Fig.~\ref{fig:dos} for different sets of parameters, both for the full and the half shell cases.  In the numerical calculation of the DOS we used 140 discrete values of the wave vector $k$, as if the nanowire had a finite length $L_z=50R_{\rm ext}$, which is sufficient to obtain DOS values close to the infinite-length limit.

To calibrate our model, we tried different values of the effective superconductor-semiconductor coupling parameter $\gamma$ and selected one that gives an induced superconducting gap comparable to the gap observed in experiments \cite{Vaitiekenas20,Stampfer22,Zellekens25}. We consider an induced gap of at least one quarter of  the original value $\Delta$ in the full shell case, i.e. about 0.08 meV, which could be obtained with $\gamma=83$ meV. We note that the induced gap also depends on the effective electrostatic potential, in particular on the parameters characterizing the QW potential. A deeper well, for example, results in the electrons being attracted more toward the superconductor-semiconductor interface, which leads to a larger induced gap. Conversely, reducing the depth the QW, or its volume (e.g., going from a full-shell to a half-shell) results in a smaller value of the induced gap.
In Fig.~\ref{fig:dos}, the solid curves correspond to $\gamma=83~$meV and three values of the chemical potential, $\mu=3.3, 5.0, 6.6$ meV, measured from the ground state.  The low DOS region (at low energies) indicates the presence of an induced gap. Note that for the full shell case the value of the induced gap has a weak dependence on the chemical potential, as the electrons are always in close proximity to the metallic superconductor and bound there by the radial QW potential.  The situation changes for the half shell case, where the electrons are in close contact with the metallic superconductor only at chemical potentials, when only low-energy transverse modes, like (b1)-(b4) shown in Fig.~\ref{fig:tm}, are populated.  For larger chemical potentials, the delocalized modes like (b5) and (b6) (or higher) become active and the induced superconductivity weakens, being characterized by induced gap values on the order of $10-20~\upmu$eV [see Fig. \ref{fig:dos}(b)].  

\blue{The effect of decreasing the effective semiconductor-superconductor coupling is illustrated by considering (full- and half-shell) hybrid systems with $\gamma=41.5~$meV, i.e., half the original value, and $\mu=3~$meV. The  corresponding DOS is represented by the dotted lines in Fig.~\ref{fig:dos}. Note the (significant) reduction of the induced gap as compared to the $\gamma=83~$meV case (full red lines). Furthermore, the effect of the QW potential is illustrated by considering hybrid systems with $\mu=3~$meV and no QW. The corresponding DOS curves -- dashed lines in Fig.~\ref{fig:dos} -- show a massive reduction of the induced gap (by a factor $\sim 5-10$), which is the result of the transverse modes having a smaller amplitude at the semiconductor-superconductor interface in the absence of the (attractive) QW potential. In this context, we point out that transverse modes located away from the interface, e.g., modes that have most of their spectral weight in the unproximitized half of the half-shell hybrid wire, will be characterized by a nearly zero induced gap and will provide normal conducting channels.}

\begin{figure}
\centering
\includegraphics[trim=100 0 140 0, clip, scale=0.62]{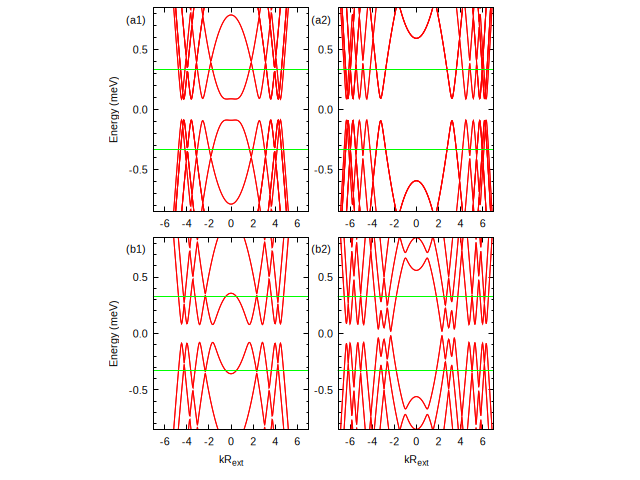}  \\[-5mm]
\caption{Low energy BdG spectra of infinite length nanowires within the static approximation. 
The full shell case is shown in panels (a1) and (a2) with $\mu = 3.3$ meV and $\mu=6.6$ meV, respectively. The half shell case is shown in panels (b1) and (b2), again with $\mu = 3.3$ and $\mu=6.6$ meV. The green horizontal lines indicate the superconducting gap of the parent superconductor, $\Delta = 0.33$ meV. Note that the static approximation is highly accurate for energies $E\lesssim 0.4\Delta\approx 0.13~$meV, i.e. near the gap edge.} 
\label{fig:ie}
\end{figure}

\blue{To better understand the physics that controls the size of the superconducting gap induced in the semiconductor, we compute the energy spectrum of the effective BdG Hamiltonian given by Eq.~(\ref{eq:heff}), i.e., the low-energy spectrum of the hybrid system in the static approximation, Eq.~(\ref{eq:ses}). Figure~\ref{fig:ie} shows the energy spectra for the full shell (a1 and a2) and half shell (b1 and b2) systems with chemical potentials $\mu=3.3$ meV (left panels) and 6.6 meV (right panels).  Note that the spectra are consistent with the DOS results shown in Fig.~\ref{fig:dos}, with the low-energy DOS maxima being associated with positive energy minima corresponding to different subbands.}

\blue{We point out that the static approximation is valid for energies (much) smaller than $\Delta$, becoming exact in the limit $E\rightarrow 0$. In practice, the eigenvalues obtained within this approximation are highly accurate up to energy values on the order of half $\Delta$ \cite{Paudel2025}.}  
However, in Fig.~\ref{fig:ie} we show a significantly larger energy interval, to clearly identify the subbands associated with different transverse modes and better understand the corresponding subband folding. Each transverse mode leads to a (typically sharp) local minimum of the effective positive energies (or maximum of the negative energies), within the energy gap of the parent superconductor, $-\Delta \leq E\leq \Delta$, as a function of the wave vector $k$.

\blue{The characteristic wave vectors associated with these minima/maxima are, practically, the Fermi wave vectors $k_F^{(n)}$ corresponding to each subband, for a given value of the chemical potential. Consequently, the high-$k$ minima/maxima in Fig.~\ref{fig:ie} are associated with low-energy subbands, while the lowest-$k$ minimum corresponds to the top occupied mode.}
Thus, we can associate to each transverse mode $n$ a specific value $\widetilde{\Delta}_n$ of the induced superconducting gap given by half the difference between the local maximum and minimum at the corresponding $k_F^{(n)}$ value.  We observe that for the full-shell case, Fig.~\ref{fig:ie} (a1) and (a2), the gaps associated with different modes are nearly equal and very weakly dependent on chemical potential.  
\blue{In the half-shell case, these properties still hold for the four lowest energy modes, i.e. the modes shown in Fig.~\ref{fig:tm}(b1-b4), which are strongly localized near the QW. These are the only occupied modes for $\mu=3.3~$meV -- see Fig.~\ref{fig:ie} (b1) -- and are responsible for the large-$k$ minima/maxima at higher values of the chemical potential seen on the left and right quarters of Fig.~\ref{fig:ie}(b2).} However, for $\mu \gtrsim 5~$meV the delocalized modes become active and generate lower values of the induced gap, due to their smaller amplitude in the vicinity of the semiconductor-superconductor interface. This is illustrated in Fig.~\ref{fig:ie}(b2), where one can clearly notice the smaller induced gap values associated with the three lowest-$k$ minima/maxima and, particularly, the induced gap characterizing the top occupied subband, which controls the overall value of the induced gap for this system, $\widetilde{\Delta}\approx 170~\upmu$eV.

\blue{The finite induced gap characterizing both the full-shell and the half-shell systems reveals the presence of (proximity-induced) superconductivity inside the semiconductor wire. A more complete characterization of the induced superconductivity includes the eigenstates of the hybrid system, which, generically, have both electron and hole components. While the states that are sufficiently far away from the gap edge have almost pure electron or hole characters (depending on whether they correspond to bands associated with the normal state spectra -- see Fig.~\ref{fig:ien} -- or to their ``flipped'' hole 
counterparts, respectively), near the gap edge the BdG states are expected to have comparable electron and hole components. The presence of tubular superconducting states having comparable electron and hole components is the necessary condition for observing Little-Parks oscillations in the core-shell nanowires.}

\begin{figure}
\centering
\includegraphics[trim=25 50 10 40, clip, scale=0.42]{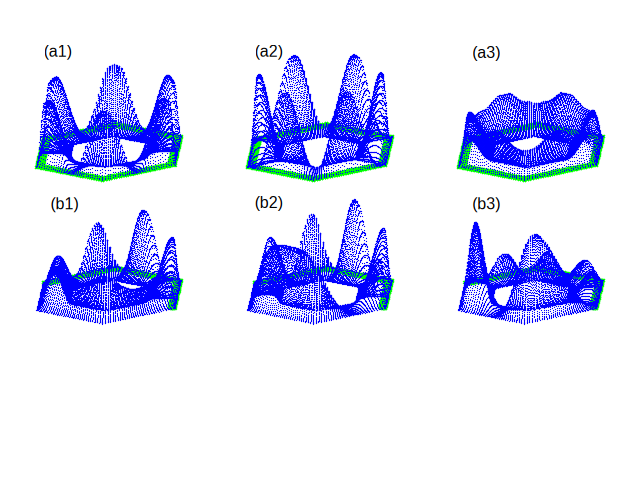} \\[-12mm]
\caption{Transverse spatial profiles of the hole components of gap-edge wave functions corresponding to the positive energy minima associated with different subbands. The corresponding electron components (not shown) have similar profiles and comparable spectral weights.
Panels (a1-a3) correspond to the full-shell case with the energy spectrum shown in Fig.~\ref{fig:ie}(a2) and wave vectors  $kR_{\rm ext}=\pm~5.2, \ \pm~4.5, \ {\rm and}\ \pm3.2$, respectively.  Panels (b1-b3) correspond to the half-shell case illustrated in Fig.~\ref{fig:ie}(b2) and wave vectors $kR_{\rm ext}=\pm~3.4, \ \pm~2.9, \ {\rm and}\ \pm~2.3$, respectively. 
}
    \label{fig:lholes}
\end{figure}

\blue{To demonstrate the presence of this type of states, we first consider the full-shell case and calculate the cross-section spatial distribution of the {\em hole} components corresponding to positive energy BdG states at the subband minima in Fig.~\ref{fig:ie}(a2) with energies $\sim 0.09$ meV and characteristic  wave vectors $kR_{\rm ext}=\pm 5.2, \ \pm 4.5, \ {\rm and}\ \pm 3.2$. The results are shown in Fig.~\ref{fig:lholes} (a1), (a2), and (a3), respectively. Note the similarity with the spatial distribution of the corresponding normal state transverse modes shown in  Fig.~\ref{fig:tm} (a4), (a5), and (a6), respectively. }

The analog gap-edge states for the half-shell system with energies 0.036 meV, 0.051 meV, and 0.017 meV and wave vectors 
$kR_{\rm ext}=\pm 3.4, \ \pm 2.9, \ {\rm and}\ \pm 2.3$, 
respectively, are characterized by hole component profiles as shown in Fig.~\ref{fig:lholes} (b1), (b2), and (b3), respectively. 
The spatial distributions of the first two states are similar to the (normal state) transverse modes represented in Fig.~\ref{fig:tm} in the panels (b5) and (b6), while the transverse profile of the top occupied (seventh) subband is not included in that figure. \blue{The smaller induced gap characterizing the top subband, Fig.~\ref{fig:lholes}(b3), as compared to the other two states, (b1) and (b2), is a consequence of this mode having more weight outside the QW (interface) region. Note that a (nearly) vanishing spectral weight at the semiconductor-superconductor interface leads to a (nearly) vanishing induced gap for the corresponding subband.}

Finally, we need to mention that the electron components these gap-edge states corresponding to Fig.~\ref{fig:lholes} have very similar profiles (not shown) and the electron-hole weights of these states are both close to 0.5.


\blue{The main conclusion of this subsection is that half-shell hybrid nanowires can support tubular superconducting states that surround the entire circumference of the nanowire and are characterized by comparable electron and hole components, similar to the superconducting states generically supported by full-shell hybrid nanowires.}



\subsection{Nanowires with finite length}

\begin{figure}
\centering
\includegraphics[trim=0 0 0 120, clip, scale=0.39]{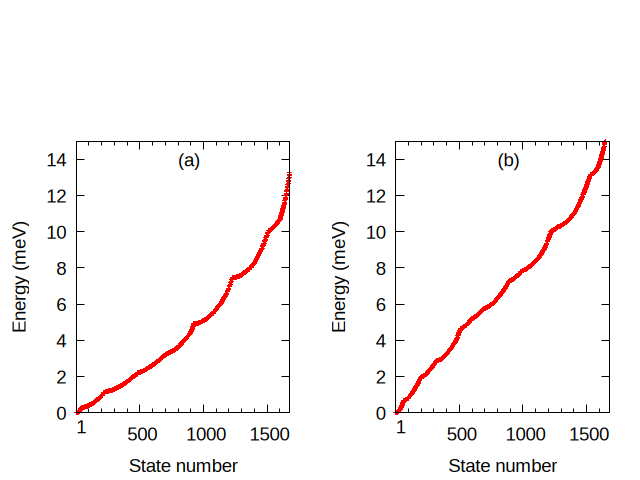}  \\[-2mm]
\caption{Energy spectra for normal state finite nanowires of length $L_z=5000~$nm with (a) full-shell and (b) half-shell. The kinks correspond to the bottoms of different transverse subband and match well the corresponding values for the infinite system shown in Fig.~\ref{fig:ie}. All states are spin degenerate.}
\label{fig:fe}
\end{figure}

To get further insight into the structure and parameter dependence of the induced superconductivity, we also performed calculations on a finite length nanowire.  For a proper comparison with the infinite length case, i.e., to minimize the finite size effects, we chose $L_z=50R_{\rm ext}=5000$ nm.  
The basis used for constructing the effective low-energy Hamiltonian contains 1680 states, including 12 radial modes, 70 longitudinal modes, and 2 spin states (see Sec. II). 
First, we diagonalize the normal state  Hamiltonian and plot the corresponding energies, sorting them in increasing order, as shown in Fig.~\ref{fig:fe}. The kinks in the spectrum correspond to the bottoms of different transverse subbands. Note that the corresponding energies match well the subband minima of the continuous spectra shown in Fig.~\ref{fig:ien}.
%

\begin{figure}
\centering
\includegraphics[width=\linewidth]{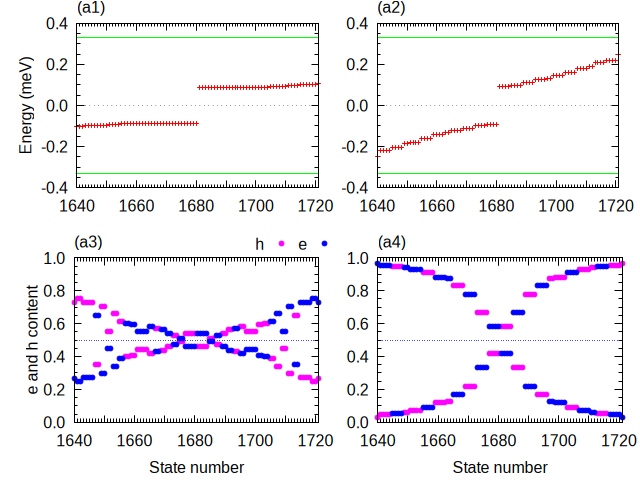}
\includegraphics[width=\linewidth]{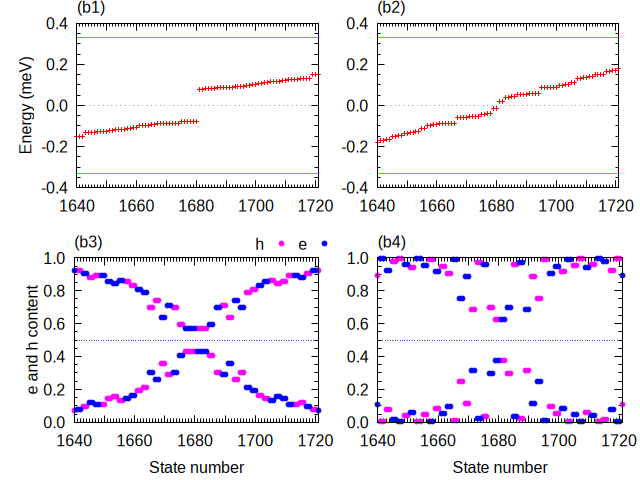} \\[-2mm]
\caption{Energy spectra and corresponding particle-hole content for a full-shell nanowire (top four panels) and a half-shell system (bottom four panels). 
In (a1) and (a2) we show the effective BdG energies for chemical potentials $\mu=3.3$ meV and $\mu=6.6$ meV, respectively. Below, in (a3) and (a4) we show the electron (e) and hole (h) content of the corresponding states. The same information is shown in the panels (b1)-(b4) for the half shell nanowire.}
    \label{fig:feff}
\end{figure}

\blue{Next, we consider the superconducting phase and focus on calculating the energy spectrum and the electron/hole content of the corresponding states.
Using the same parameter values as for the infinite wires, we solve the effective eigenstate equation, Eq.~(\ref{eq:heff}), for a finite  wire  of length $L_z$.
The results for both full- and half-shell nanowires and two different values of the chemical potential are shown in Fig.~\ref{fig:feff}.}

The transition between negative and positive energies occurs between states number 1680 and 1681. This time we restrict the energies to values compatible to the static approximation, reasonably smaller than $\Delta$ (indicated by the green horizontal line).  As before, for the full shell case the induced gap does not change much when the chemical potential increases, from 3.3 meV to 6.6 meV, but only the energy dispersion does, as shown in  Figs.~\ref{fig:feff}(a1) and (a2). Again, these figures can be seen as another representation, i.e. the discrete version, of the effective spectra obtained for the infinite length.  The energies can be 2-fold or 4-fold  degenerated (spin only, or both spin and orbital), depending on the transverse modes involved.

In Figs.~\ref{fig:feff} (a3) and (a4) we show the electron ($\eta=1$) and hole ($\eta=-1$) components of the effective states, corresponding to the energies presented above, in panels (a1) and (a2), respectively. The electron and hole components of the same state are comparable in the middle of the spectra, i.e. at the edges of the superconductivity gap, indicating a strong mixing in the Nambu space.  The mixing decreases outside this region, more or less steadily, depending on the chemical potential.  A low chemical potential selects the transverse modes tightly connected to the metallic superconductor, and a higher chemical potential selects modes which are more loosely distributed in the radial direction. In addition, discontinuities of this mixing as seen for states 1697-1700 or 1709-1710 in example (a3) are associated with the changes of the radial or longitudinal modes at those particular energies.   

Figs.~\ref{fig:feff}(b1)-(b4) show the analog results for the half shell case.  The lower chemical potential activates only the modes bound to the QW (shown in Fig.~\ref{fig:tm}(b1)-(b4)) and the situation represented in Fig.~\ref{fig:feff}(b3) is almost the same as for the full shell.  Instead, for the higher chemical potential, the induced superconducting gap is very small, comparable to the tiny gaps between the normal discrete energy states due to the finite length of the nanowire. Nevertheless,  although the gap is so small, some effective states remain superconducting.  In Fig.~\ref{fig:feff}(b4) we identify these states as having a strong or considerable electron-hole mixing, on a background of nearly normal states, which have the electron or hole content nearly 1 or nearly 0.  The superconducting states now are distributed in the entire volume of the semiconductor material, since they involve the delocalized radial modes, like those shown in Fig.~\ref{fig:tm}(b5)-(b6), or modes with higher energies. Each radial mode of the normal state generates a subset of effective superconducting states. Within each subset the superconducting character vanishes monotonically with increasing the energy, i.e. the hole content (hc) approaches 0 or 1, but different subsets are intercalated.  For example, the 5-th radial mode shown in Fig.~\ref{fig:tm}(b5) yields the (spin degenerated) states 1683-1684 (hc 0.30), 1691-1692 (hc 0.89), 1699-1700 (hc 0.05), 1705-1706 (hc 0.96), and 1715-1716 (hc 0.02).  While the 6-th radial mode, Fig.~\ref{fig:tm}(b6), gives states 1689-1690 (hc 0.31), 1693-1694 (hc 0.75), 1697-1698 (hc 0.10), 1701-1702 (hc 0.92), 1709-1710 (hc 0.06), 1711-1712 (hc 0.96).  A slight mixing of the radial modes can also occur because of the coupling of the semiconductor and superconductor via the self energy, Eq.~(\ref{eq:se}).

\begin{figure}
\centering
\includegraphics[trim=0 0 0 100, clip, scale=0.39]{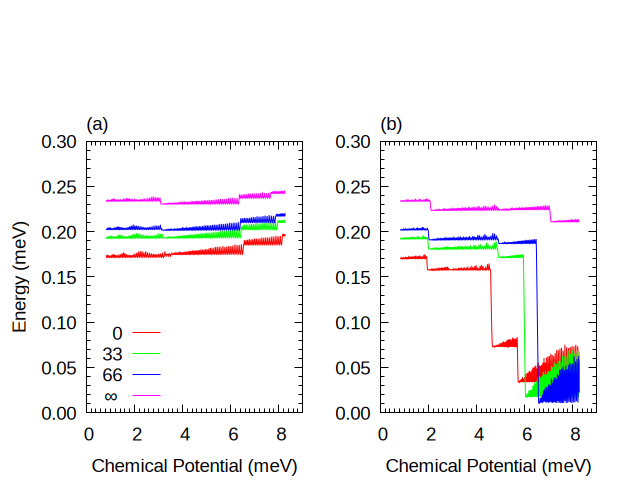}  
\caption{The effective superconducting gap of (a) the full shell model and (b) the half shell model, as function of the chemical potential measured relatively to the ground state of the normal nanowire.  The width of the semiconductor is 40 nm and a rectangular potential barrier of 10 nm width is placed at the inner boundary. The red lines are obtained without this barrier. The infinite barrier was simulated using a height of 8300 meV. }
    \label{fig:thick}
\end{figure}

In order to further investigate the effect of radial localization of the transverse modes on the superconductivity induced in the tubular semiconductor we studied the evolution of the induced superconductivity gap as function of chemical potential, which is shown in Fig.~\ref{fig:thick}.   This gap is now defined as the energy interval between the first positive and last negative energies, which correspond to the states 1681 and 1680, and readable in Fig.~\ref{fig:feff}~(a1)-(a2) and (b1)-(b2). (Or twice the absolute value of either these energies, because of the electron-hole symmetry.)

In this series of calculations we also considered the effect of a band-offset potential at the interface between the semiconductor shell and the semiconductor core \cite{Sitek25b}.  We modeled this band-offset with a thin, 10 nm wide, rectangular potential barrier at the inner boundary of the semiconductor shell.  
Increasing the barrier height from zero (no barrier) to 33 and then to 66 meV results in pushing the wave functions towards the superconducting shell and effectively narrowing the semiconducting shell which becomes 30-nm thick for the infinite barrier. At the same time the wave functions get in tighter contact with the metallic semiconductor, and consequently the superconductivity gap increases, both in the full and half shell cases, as a function of the barrier height. 

With a fixed barrier height (i.e. fixed effective width of the semiconductor shell), the induced superconductivity gap varies in steps when the chemical potential increases. We notice slight positive steps in the full-shell case, Fig.~\ref{fig:thick}(a).  The chemical potential values at the steps roughly correspond to the low density of states of the normal spectrum shown in Fig.~\ref{fig:fe}(a).  In the half-shell case, the steps are created by new transverse modes activated in the superconductivity window. The transverse modes become more distributed when the energy increases, the contact with the parent superconductor weakens, and the induced gap have negative, decreasing steps, as seen in Fig.~\ref{fig:thick}(b). We also see how the gap collapses to low or very low values at large chemical potentials, when the delocalized transverse modes become populated. This transition point depends on the effective thickness of the semiconductor shell, controlled by the height of the potential barrier at the core side, such that the induced gap is more robust for a thinner shell.  However, it is interesting to notice that when the delocalized transverse modes become active, i.e. for chemical potentials above 5.8 meV, the (small) induced gap becomes smaller for a thinner shell.  This reversed gap order occurs because the transition between the transverse modes localized in the QW potential and the modes expanded over the entire hexagonal shell is more gradual in a thicker shell and more abrupt in a thinner one.    

\begin{figure}
\centering
\includegraphics[trim=0 0 0 100, clip, scale=0.39]{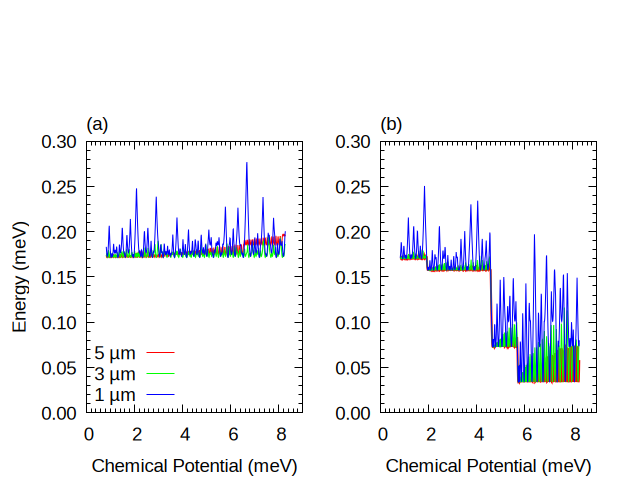}  
\caption{The effective superconducting gap of (a) the full shell model and (b) the half shell model, as a function of the chemical potential measured relatively to the ground state of the normal nanowire, and for different nanowire lengths $L_z$.  The red line is the same as in Fig.~\ref{fig:thick} obtained for $L_z=5000$ nm. The green and blue lines correspond to $L_z=3000$ nm and $Lz=1000$ nm, respectively.
}
    \label{fig:length}
\end{figure}

The dense oscillations of the superconducting gap observable in Fig.~\ref{fig:thick}  are caused by the multiple longitudinal modes present at the chemical potential level.  Such oscillations can be better observed for shorter nanowires, as we show in Fig.~\ref{fig:length}. Now the thickness of the semiconductor shell is kept constant (40 nm with zero potential barrier) while the length $L_z$ is varied. The red line is the same as in Fig.~\ref{fig:thick} ($L_z=$ 5000 nm), for comparison. Reducing the length of the wire results in the increase of the amplitude and decrease of the frequency of the oscillating superconducting gap because of sparser discrete longitudinal modes.  These oscillations reflect the discrete nature of the longitudinal modes and - remarkably - in the half shell case their amplitude can be comparable, or even larger, than in the full shell case.  The minimum values of the induced gap remain more stable than the maxima, and correspond to the situations when the chemical potential is near a discrete energy. 
The maxima of the induced gap correspond to chemical potentials situated in between discrete states, in the gaps created by the quantum confinement, or in other words, due to the finite size effects.

\section{Conclusions \label{sec:conc}}

We have investigated the induced superconductivity within a 
\blue{core-shell} semiconductor nanowire covered by a 
\blue{superconducting film.} The main motivation was the recent experimental results showing that superconductivity propagates within the entire 
semiconductor \blue{shell}, irrespective of whether the 
\blue{superconducting film covers the wire entirely, forming 
a full metallic shell \cite{Vaitiekenas20}, or only partially, e.g., as a half metallic shell \cite{Stampfer22,Zellekens25}.}

We use an effective Hamiltonian that naturally incorporates the coupling between the superconductor and the semiconductor subsystems,  which can be derived from the Green's function in the low energy limit, providing a robust method for evaluating both the full-shell and half-shell configurations. \blue{The effective electrostatic potential generated by the work function difference between the semiconductor and the superconductor is modeled as a narrow potential well near the semiconductor-superconductor interface.}
Our results demonstrate that the presence of the quantum well at the interface drastically affects \blue{the transverse profiles of the semiconductor subbands and, in turn,}  the induced superconductivity 
associated to \blue{different modes, which increases monotonically with the interface spectral weight of the corresponding mode.}
\blue{Consequently, in a full-shell system the induced gap has similar values for different subbands and is weakly dependent on the chemical potential. By contrast, in a half-shell system the subband and chemical potential dependence of the induced gap is strong, with typical values that are significantly smaller than those characterizing their full-shell counterparts. In addition, the states of the full-shell system are always tubular, surrounding the entire circumference of the nanowire. By contrast, the lowest energy states of the half-shell system are strongly localized near the quantum well (interface) region and only the modes with high-enough energy are tubular.}

\blue{After characterizing the spatial profiles of the low energy modes and establishing the presence of the tubular-type states in both full-shell and half-shell systems}, we identify the 
superconducting quantum states that 
actually can carry the supercurrent observed in experiments. 
To determine whether a state is normal or superconducting we examined the electron and hole components of the eigenvalues of the effective BdG Hamiltonian. A normal state has only one dominant component, whereas a superconducting state is a combination of electron and hole components, with comparable contributions from both.
For the full-shell coverage, \blue{as the induced gap is about the same for all transverse subbands resulting in a relatively well-defined gap edge, } 
the superconducting character of individual states 
\blue{decreases} monotonically when the \blue{absolute values of the} effective excitation energy increases \blue{away from the gap edge.  
On the other hand, in the half-shell system with high-enough chemical potential,} 
the electron-hole mixing has a non-monotonic energy dependence, with some higher excited states states being more superconductive than some lower states, which may exhibit a quasi normal behavior. The reason for this behavior is 
the presence of two types of transverse modes, some localized within the semiconductor-metal quantum well, others extended within the entire volume of the semiconductor \blue{shell}. 

Each transverse mode generates a subset of effective states with different superconducting character.  In this case, the classical concept of coherence length has no longer a global meaning, but 
\blue{should be applied to specific subbands, i.e. transverse modes.}
Consequently, in the full-shell case superconductivity is manifestly induced over the entire circumference of the semiconductor nanowire. \blue{In the half-shell case, on the other hand, the lowest energy modes are strongly localized by the effective electrostatic potential near the semiconductor-superconductor interface (i.e., the QW region), while higher energy modes are delocalized throughout the entire semiconductor shell, having tubular profiles.
Thus, a key requirement for obtaining tubular superconducting states, which can explain the observed Little-Parks effect  \cite{Stampfer22,Zellekens25}, is that the chemical potential is sufficiently high.}

\blue{The core-shell structure of the semiconductor nanowire is critical for obtaining tubular superconducting states. We show that} 
the induced superconducting gap increases if the thickness of the semiconductor shell decreases,    
\blue{as a results of enhancing the spectral weight at the semiconductor-superconductor interface}, for both the full and the half shell.

\blue{In the half-shell case, we find that the presence of subband-dependent induced gaps is revealed 
by a stepwise dependence of the quasiparticle gap on the chemical potential. The transition between two successive steps is associated with a new subband being occupied. The step values decrease strongly when the delocalized modes having significant weight away from the semiconductor-superconductor interface become active. Note that the fluctuations of the gap values superimposed on the stepwise dependence on the chemical potential is a finite size effect that vanishes in the long wire limit.} 
 
\blue{Finally, we point out two important aspects that were not explicitly addressed in this work. First, there is a natural question concerning how much of the semiconductor wire surface has to be proximitized by the superconducting film to obtain tubular induced superconductivity, i.e., within the entire semiconductor shell.} 
Our calculations suggest that \blue{covering only two, or even one facet of the wire may be sufficient,}
if the device parameters are properly chosen. Experimental detection of Little–Parks oscillations in such a device would be illuminating.
\blue{Second, one may be concerned that disorder could destroy the coherence effects discussed here. Indeed, disorder tends to induce localization, in particular radial localization that may affect the tubular nature of the superconducting states. While the quantitative details have to be investigated explicitly, we point out that states with high-enough values of the wave vector and angular momentum are delocalized even in the presence of disorder. Hence, we expect the physics described in this work to hold in the presence of disorder in systems with high-enough chemical potential.}

\begin{acknowledgments}
This work was financed by Reykjavik University Research Fund, Grant 223016. T.D.S. acknowledges 
ONR-N00014231206.
We are  thankful to Patrick Zellekens, Russell Deacon, Farah Basaric, and Thomas Sch\"apers for detailed discussions on their experimental work \cite{Zellekens25}. 
\end{acknowledgments}

\bibliographystyle{apsrev4-2}

\bibliography{csmaj}

\end{document}